\documentclass{kapproc}

  \kluwerbib

\usepackage{graphicx}

\begin{document}

\articletitle{GAIA:~~AGB stars as tracers of \\star formation
histories in \\the Galaxy and beyond}

\author{Ar\={u}nas Ku\v{c}inskas}
\affil{Lund Observatory, Box 43, SE-221 00, Lund, Sweden}
\affil{Institute of Theoretical Physics and Astronomy,
Go\v{s}tauto 12, Vilnius 2600, Lithuania}
\vskip -20pt

\author{Lennart Lindegren}
\affil{Lund Observatory, Box 43, SE-221 00, Lund, Sweden}
\vskip -20pt

\author{Toshihiko Tanab\'{e}}
\affil{Institute of Astronomy, The University of Tokyo, Tokyo,
181-0015, Japan}
\vskip -20pt

\author{Vladas Vansevi\v{c}ius}
\affil{Institute of Physics, Go\v{s}tauto 12, Vilnius 2600,
Lithuania}
\vskip -10pt

\chaptitlerunninghead{GAIA: AGB stars as tracers of star formation
histories}

\begin{abstract}

\noindent We discuss the tracing of star formation histories with
ESA's space astrometry mission GAIA, emphasizing the advantages of
AGB stars for this purpose. GAIA's microarcsecond-level astrometry,
multi-band photometry and spectroscopy will provide individual
distances, motions, $T_{\rm eff}$, $\log g$ and $[{\rm M}/{\rm
H}]$ for vast numbers of AGB stars in the Galaxy and beyond.
Reliable ages of AGB stars can be determined to distances of
$\sim$200 kpc in a wide range of ages and metallicities, allowing
star formation histories to be studied in a diversity of
astrophysical environments.

\end{abstract}


\section{The GAIA mission, an astrometric and spectrophotometric
survey of the Galaxy}

\noindent The European Space Agency's GAIA mission, approved for
launch in 2010--12, aims at surveying the Galaxy to 20th visual
magnitude, using a combination of astrometric measurements (for
trigonometric parallaxes and proper motions), multiband photometry
(for basic stellar parameters like temperature and metallicity),
and radial-velocity measurements. Targeted accuracies versus
magnitude allow direct distances and motions to be obtained for
large samples of intrinsically bright stars across the Galaxy and
in some nearby Local Group galaxies. Expected typical accuracies
are shown in Table~1. In total more than 1~billion stars will be
observed, of which 50--100~million will obtain individual parallax
distances to better than 5~per cent. A primary science goal is to
study formation, evolution and structure of the Galaxy, for which
large-scale mappings of star formation histories are essential.
For a full description of the very broad range of science goals
see Perryman et al. (2001).

In its present design GAIA comprises two astrometric instruments,
with $1.4\times 0.5$~m$^2$ apertures and a combined 0.5~deg$^2$
field of view, and a separate photometric/spectroscopic instrument
with a $0.5\times 0.5$~m$^2$ aperture. The latter performs
photometry in $\sim\,$11 bands for astrophysical classification,
and $R\sim 10^4$ spectroscopy in the 849--874~nm wavelength range,
mainly for radial velocities. During its lifetime of at least
5~years, the satellite will scan the entire sky repeatedly, so
that each object is observed at multiple epochs. The above numbers
and accuracy predictions refer to the recently (May 2002)
completed revised design, aiming at a substantially reduced
mission cost compared with the previous baseline (Perryman et al.
2001), while preserving all science goals intact.

\begin{table}
\caption[]{ Predicted accuracies, versus $V$ magnitude, of
individual AGB stars observed by GAIA. Standard errors estimated
by Lindegren (unpublished, 2002) for parallax ($\pi$) and proper
motion ($\mu$), by Katz \& Munari (2002) for radial velocity
($v_r$), and by Vansevi\v{c}ius et al.\ (2002) for photometrically
derived $T_{\rm eff}$, $\log g$ and $[{\rm M}/{\rm H}]$. The last
column is the maximum distance of an AGB star ($M_V \simeq -2$) at
the given accuracies. }
\begin{center}
\begin{tabular}{cccccccc}
\sphline $V$ & $\sigma(\pi)$ & $\sigma(\mu)$ & $\sigma(v_r)$ &
$\sigma(\log T_{\rm eff})$ & $\sigma(\log g)$ &
$\sigma([{\rm M}/{\rm H}])$ & $d_{\rm max}$ \\
mag & $\mu$as & $\mu$as~yr$^{-1}$ & km~s$^{-1}$ & &&& kpc \\
\sphline 15 & \phantom{0}13 & \phantom{0}8 & 1.1 & 0.007
& 0.20           & 0.24
       & \phantom{0}25 \\
17 & \phantom{0}32 &           18 & 6.3 & 0.01\phantom{0} & 0.27
& 0.32
       & \phantom{0}60 \\
19 & \phantom{0}90 &           50 &  -- & 0.04\phantom{0} & 0.60
& 0.63
       &           150 \\
20 &           160 &           90 &  -- & 0.13\phantom{0} &
1.1\phantom{0} &
1.3\phantom{0} &           250 \\
\sphline
\end{tabular}
\end{center}
\end{table}

\section{Tracing stellar populations using AGB stars}

\noindent
The availability of precise photometry is essential for age
derivations using isochrone fitting to the main sequence turn-off
(MSTO) point. Simulations of GAIA photometry demonstrate that this
method may be successfully exploited with GAIA even in such
distant stellar systems as the Magellanic Clouds (Ku\v{c}inskas et
al.\ 2002), but only for populations younger than $\sim 1$~Gyr.
In this paper we argue that GAIA observations of AGB stars can be
used to determine star formation histories to even greater distances
and for much older populations.

GAIA will provide a wealth of astrometric and
spectrophotometric data on galactic and extragalactic AGB stars.
Their uses are at least twofold: (a) as kinematic tracers, using
distances and space motions obtained from the astrometric and
radial-velocity data; (b) for age determinations, using basic
stellar-atmosphere parameters ($T_{\rm eff}$, $\log g$ and
$[{\rm M}/{\rm H}]$) derived from the spectrophotometric data,
combined with distances and theoretical isochrones.

From the astrometric and radial-velocity accuracies in Table~1 it
is obvious that GAIA will yield accurate distances ($<10$\%) and
full space velocities ($<1$~km~s$^{-1}$) for individual AGB stars
up to distances of $10$--$15$~kpc, if no interstellar extinction
is present.

Extensive simulations by the Vilnius GAIA group (Vansevi\v{c}ius
et al.\ 2002; Ku\v{c}inskas et al.\ 2002) show that GAIA will also
provide precise metallicities ($\sigma([{\rm M}/{\rm H}]) \leq
0.3$) and gravities ($\sigma(\log g) \leq 0.3$) for AGB stars
brighter than $V \sim 17$ (Table~1). Precise effective
temperatures ($\sigma(\log T_{\rm eff}) \leq 0.04$) are derived
down to $V \sim 19$. This holds within a broad range of
metallicities ($[{\rm M}/{\rm H}]>-2$) and ages (0.05--15~Gyr).

Metallicity estimates of intermediate age and old stellar
populations can also be obtained from the slope of the red giant
branch (e.g.\ Ferraro et al.\ 2000). Our simulations show that the
method could provide an independent estimate of $[{\rm
M}/{\rm H}]$ with GAIA, effective up to distances of $\sim
200$~kpc, if no interstellar extinction is present (Ku\v{c}inskas
et al.\ 2002).

We have recently shown (Ku\v{c}inskas et al.\ 2000) that reliable
ages can be derived using isochrone fits
to the AGB sequences on the observed HR diagram. It is essential
for this procedure to have precise effective temperatures of the
AGB stars, which can be derived by fitting synthetic spectral
energy distributions to observed photometric fluxes (e.g.,
{\it BVRIJHK\/}). The method was successfully tested and compared
with the MSTO method on a sample of populous star clusters in the
Magellanic Clouds spanning a wide range of ages (Table~2 and Fig.~1).
For galactic AGB stars, it is clear that the distance
information needed to construct the observational HR diagrams will
be available through GAIA. It thus appears that precise age
estimates ($\sigma (\log t)<0.3$) can be obtained for a wide range
of ages (0.05--10~Gyr) and metallicities ($[{\rm M}/{\rm H}]>-2$).

\section{Conclusions}

\noindent GAIA will provide unique astrometric and photometric
data for studying individual and collective properties of stars in
the Galaxy and its surroundings. AGB stars, being intrinsically
bright, will provide precise individual distances, kinematics,
$T_{\rm eff}$, $\log g$ and $[{\rm M}/{\rm H}]$ up to distances of
$10$--15~kpc. Using isochrone fitting to the AGB stars will give
reliable ages ($\sigma(\log t)<0.3$) for a wide range of ages and
metallicities. If distances are known by other means (e.g.\ in
distant clusters), the method can be used up to $\sim 200$~kpc.
Thus, AGB stars will allow the formation histories and kinematics
of stellar populations to be probed in a diversity of astrophysical
environments both in the Milky Way and in neighbouring galaxies.


\begin{acknowledgments}
\noindent The work was supported by a grant (NB00-NO30) of the
Nordic Council of Ministers and by a grant of the Wenner-Gren
Foundations. AK thanks the Workshop organisers for financial
support to attend the event.
\end{acknowledgments}

\begin{table}[t]
\parbox{6.3cm}
{ 
\caption[]{MSTO and AGB ages for a sample of LMC
and SMC clusters. AGB ages (Ku\v{c}inskas et al., in preparation)
were derived using the same cluster metallicities (Col.~2) as for
the MSTO estimates.}
\begin{tabular}{lccc}
\sphline
Cluster     &  [Fe/H]  &      MSTO     &      AGB      \\
\sphline
LMC:        &          &               &               \\
~~NGC 1783  &  $-0.4$  &  $0.9\pm0.4$  &  $0.8\pm0.2$  \\
~~NGC 1846  &  $-0.7$  &       --      &  $1.3\pm0.3$  \\
~~NGC 1978  &  $-0.7$  &  $2.0\pm0.2$  &  $1.5\pm0.5$  \\
~~NGC 1987  &  $-0.4$  &       --      &  $1.0\pm0.2$  \\
~~NGC 2121  &  $-0.7$  &  $3.2\pm0.5$  &  $3.5\pm0.5$  \\
SMC:        &          &               &               \\
~~Kron 3    &  $-1.3$  &  $8.0\pm0.3$  &  $9\pm3$      \\
~~NGC 152   &  $-1.0$  &  $0.8$        &  $1.2$        \\
~~NGC 419   &  $-0.7$  &  $1.2\pm0.5$  &  $1.4\pm0.2$  \\
\sphline
\end{tabular}
}
\end{table}

\begin{figure}[t]
\vskip -7.8cm \hskip 7.0cm
\parbox{5.0cm}
{\centerline{\includegraphics[height=5.5cm]{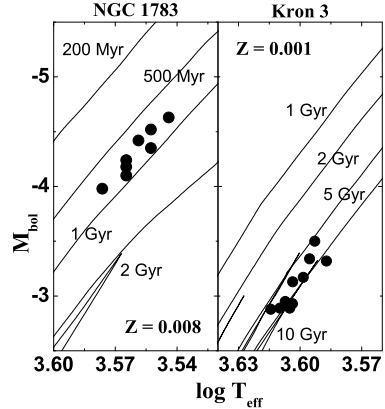}}
\vskip -1mm \caption{AGB sequences in NGC 1783 and Kron 3.
Isochrones are from Bertelli et al.\ (1994).}}
\end{figure}



\begin{chapthebibliography}{1}

\bibitem[]{B94} Bertelli, G., Bressan, A., Chiosi, C., Fagotto, F.,
    Nasi, E.: 1994, {\em A\&AS}, {\bf 106}, 275.
\bibitem[]{F00} Ferraro, F.R., Montegriffo, P., Origlia, L., Fusi Pecci,
    F.: 2000, {\em AJ}, {\bf 119}, 1282.
\bibitem[]{KM02} Katz, D., Munari, U.: 2002, {\em GAIA RVS Status
    Report, RVS-CoCo-004}\\
    (\texttt{http://wwwhip.obspm.fr/gaia/rvs/bibliography/RVS-CoCo-004.txt}).
\bibitem[]{KVST00} Ku\v{c}inskas, A., Vansevi\v{c}ius, V., Sauvage,
    M., Tanab\'{e}, T.: 2000, in: Mid- and Far-Infrared Astronomy
    and Future Space Missions, eds. T, Matsumoto \& H. Shibai,
    {\em ISAS Report SP}, No. {\bf 14}, 51.
\bibitem[]{KBV02a} Ku\v{c}inskas, A., Brid\v{z}ius, A., Vansevi\v{c}ius,
    V.: 2002, {\em Ap\&SS}, {\bf 280}, 159.
\bibitem[]{P01} Perryman, M. A. C., de Boer, K. S., Gilmore, G.,
    Hog, E., Lattanzi, M. G., Lindegren, L., Luri, X., Mignard, F.,
    Pace, O., de Zeeuw, P. T., 2001, {\em A\&A}, {\bf 369}, 339.
\bibitem[]{VBD02} Vansevi\v{c}ius, V., Brid\v{z}ius, A., Drazdys,
    R.: 2002, {\em Ap\&SS}, {\bf 280}, 31.

\end{chapthebibliography}
\end{document}